\documentclass{article} 
\usepackage{geometry}

\usepackage{geometry}
\usepackage[export]{adjustbox}
 \usepackage{amsfonts}
\usepackage{amsmath}
\usepackage{authblk}
\usepackage{booktabs}
\usepackage{graphicx}
\usepackage{url}
\usepackage[margin=0.28in]{caption}
\usepackage{natbib}
\usepackage{multirow}
 \newcounter{thm}
 \newcounter{ex}
 \newcounter{re}
\DeclareMathOperator{\EX}{\mathbb{E}}
\DeclareMathOperator{\mean}{mean}
\DeclareMathOperator{\median}{median}

\title{Testing for polytomies in phylogenetic species trees using quartet frequencies}
\providecommand{\keywords}[1]{\textbf{\textit{Index terms---}} #1}

\author[1]{Erfan Sayyari \thanks{esayyari@ucsd.edu}}
\author[1]{Siavash Mirarab\thanks{smirarab@ucsd.edu}}
\affil[1]{Department of Electrical and Computer Engineering, University of California at San Diego, 9500 Gilman Drive, La Jolla, CA 92093.}

\begin{document}
\maketitle



\abstract{Phylogenetic species trees typically represent the speciation history as a bifurcating tree. 
Speciation events that simultaneously create more than two descendants, thereby creating {\em polytomies} in the phylogeny, are possible.
Moreover, the inability to resolve relationships is often shown as a (soft) polytomy.
Both types of polytomies have been traditionally studied in the context of  gene tree reconstruction from sequence data. 
However, polytomies in the species tree cannot be detected or ruled out without considering gene tree discordance.
In this paper, we describe a statistical test based on properties of the multi-species coalescent model to test the null hypothesis that a branch in an estimated species tree should be replaced by a polytomy.
On both simulated and biological datasets, we show that the null hypothesis is rejected for all but the shortest branches, and in most cases, it is retained for true polytomies. 
The test,  available as part of the ASTRAL package, can help systematists decide whether their datasets are sufficient to resolve specific relationships of interest.}


\section{Introduction}
\begin{figure}[!htb]
\centering
\includegraphics[width=.75\textwidth]{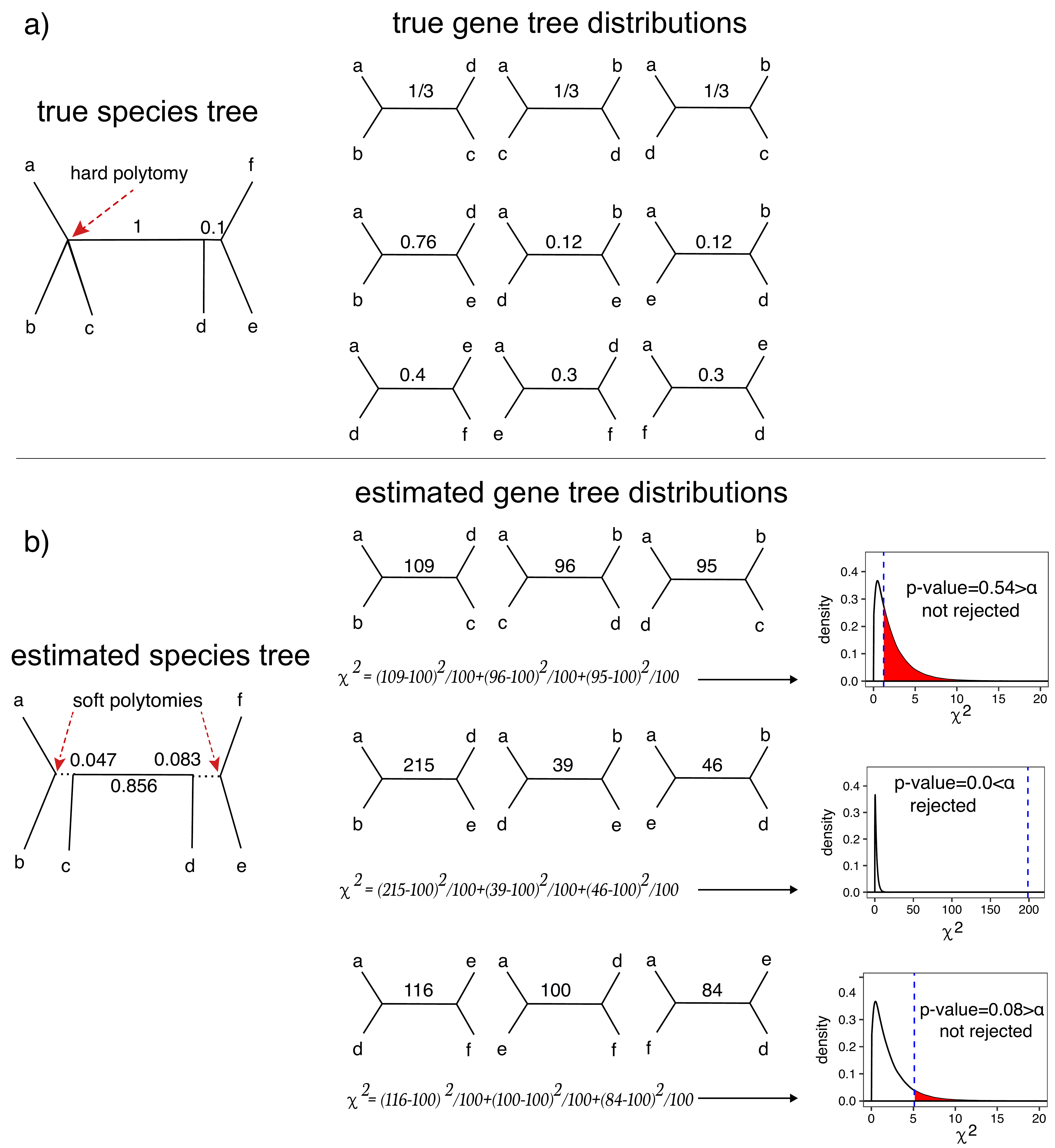}
\caption{{\bf A statistical test of polytomies.} (a) We show an example true species tree with a hard polytomy and two branches with CU lengths 1 and 0.1 (left), and  the expected quartet gene tree fractions for three selected quartets based on the MSC model (right). The first quartet is around a true (hard) polytomy and has $\frac{1}{3}$ fraction expected for all three alternative topologies. 
(b) An estimated species tree with estimated CU branch lengths (left) and a hypothetical set of quartet tree frequencies counted from 300 hypothetical gene trees (right). 
Below each set of quartets, we show the computation of the  $\chi^2$ test statistic, 
 and show where it falls on the $\chi^2$ distribution with $DF=2$; the vertical blue line shows the computed $\chi^2$ based on given counts and the shaded red areas inside distributions show the area under the $\chi^2$ distribution corresponding to the $p$-value. 
 The null hypothesis is not rejected for the branch corresponding to the true polytomy (a true negative) or the short 0.1 CU branch (a false negative);
 these branches (dotted lines in the species tree) can be replaced with soft polytomies. }
\label{fig:diagram}
\end{figure}

Phylogenies
are typically modeled as bifurcating trees. 
Even when the evolution is fully
vertical, which it is not always~\cite{Bapteste2013,Nakhleh2011},
the binary model precludes
 the possibility of several species 
evolving  simultaneously from a progenitor species~\cite{Hoelzer1994}. 
These events could be modeled in a multifurcating tree where some nodes, called polytomies, have more than two children.
True polytomies have been suggested for several parts of the tree-of-life (e.g.,~\cite{Suh2016a,Arntzen2007}).
Polytomies are also used when the analyst is unsure about some relationships due to a lack of signal in the data to resolve relationships~\cite{Townsend2012}.
The terms {\em hard} and {\em soft}
polytomies are used to distinguish between
these two cases~\cite{Maddison1989},
with a {\em soft} polytomy reserved for the case where relationships are unresolved in an estimated tree 
and a {\em hard} polytomy for multifurcations in the true tree (Fig.~\ref{fig:diagram}). 
Distinguishing the two types of polytomies is not easy. 
Moreover, the distinction between soft and hard polytomies can be blurred. 
The difficulty in resolving relationships increases as  branches become shorter.
In the limit, a branch of length zero is equivalent to a hard polytomy, which is not just difficult but impossible to resolve. 
Regardless of abstract distinctions, a major difficulty faced by systematists is to detect whether specific resolutions in their inferred trees are sufficiently supported by data to rule out a polytomy~(e.g., see~\cite{Chojnowski2008,Suh2016a}).

For any branch of a given species tree, we can pose a {\em null} hypothesis that the length of that branch is zero, and thus, the branch should be removed to create a polytomy. 
Using observed data, we can try to reject this null hypothesis, and 
if we fail to reject, we can replace the branch with a polytomy. 
The resulting polytomy is best understood as a soft polytomy because the inability to reject a null hypothesis is never accepting the alternative hypothesis. 
The inability to reject may be caused by a real (i.e., hard) polytomy, but it may also simply be due to the lack of power (Fig.~\ref{fig:diagram}).
In this paper, we present a new test with polytomy as the null hypothesis for multi-locus datasets. 

The idea of testing a polytomy as the null hypothesis and rejecting it using data has been applied to single-locus data~\cite{Jackman1999,Walsh1999}.
Likelihood ratio tests against a zero-length branch 
have been developed (e.g., SOWH test~\cite{SOWH}) and 
are implemented in popular packages such as PAUP*~\cite{paup}.
Treating polytomies as the null hypothesis  has been pioneered by Walsh {\em et al.} who sought to not only test
for polytomies but also to use a power analysis to distinguish soft and hard polytomies~\cite{Walsh1999}. 
Appraising their general framework, Braun and Kimball~\cite{Braun2001} showed that the power analysis can be sensitive to model complexity (or lack thereof).
Perhaps most relevant to our work,
Anisimova {\em et al.} presented an approximate but fast likelihood ratio test for a polytomy null hypothesis~\cite{Anisimova2006};
their test, like what we will present looks at each branch and its surrounding branches while ignoring the rest of the tree. 

The existing tests of polytomy as the null and also Bayesian methods of modeling polytomies~\cite{Lewis2005} assume that the sequence data follow a single tree.
Therefore, these methods 
test whether a {\em gene tree} includes a 
polytomy~\cite{Slowinski2001}. 
However, 
the species tree can be different from gene trees
and the discordance can have several causes, including gene duplication and loss, lateral gene transfer, and incomplete lineage sorting~\cite{degnan2009,Maddison1997}.
Arguably, the question of interest is whether the species tree includes  polytomies. 
Moreover, we are often interested to know whether we should treat
the relationship between species as unresolved given the amount of data at hand. 
These questions cannot be answered without considering gene tree discordance, an observation made previously by others as well~\cite{Slowinski2001,poe2004}.
For example, 
Poe and Chubb~\cite{poe2004}, in analyzing an avian dataset with five genes, first looked for zero-length branches in the gene trees using the SOWH test~\cite{SOWH} and found evidence that some gene trees may include polytomies. 
But they also tested if the pairwise similarity between gene trees was { greater} than a set of random trees and it was not. 
Their test of gene tree congruence, however, was not with respect to any particular model of gene tree evolution. 

A major cause of gene tree discordance is a population-level process called incomplete lineage sorting (ILS), which has been
modeled by the multi-species coalescent (MSC) model~\cite{Rannalla2003,Pamilo1988}.
The model tells us that  that likelihood of ILS causing discordance increases as branches become shorter; therefore, any test of polytomies should also consider ILS. 
The MSC model has been extensively used
for reconstructing species trees using many approaches, including Bayesian co-estimation of gene trees and species trees~\cite{Heled2010,best} and site-based approaches~\cite{SVDquartets,snapp}.
A popular approach (due to its scalability) is the summary method, where we first reconstruct gene trees individually and then summarize them to build the species tree. 
Many approaches that model ILS rely on 
dividing the dataset into quartets of
species. 
These quartet-based methods (e.g., the summary method ASTRAL~\cite{astral,astral2,astral3}, the site-based method 
SVDQuartets~\cite{SVDquartets}, and a hybrid method called Bucky-Quartet~\cite{Larget2010}) rely on the fact that 
for a quartet of species, where only three unrooted tree topologies are possible,  the species tree topology has the highest probability
of occurring in unrooted gene trees under the MSC model (Fig.~\ref{fig:diagram}a). 

Relying on the known distribution of quartet
frequencies under the MSC model, 
we previously introduced a way of computing
the support of a branch using a measure
called
 local posterior probability (localPP)~\cite{localpp-sm}. 
  In this paper, we further extend the approach used
 to compute localPP  
 to develop a fast test for the null hypothesis that a branch has length zero. 
 Under the null hypothesis, we expect that
the three unrooted quartet topologies defined around the branch should have equal frequencies~\cite{Allman}. 
 This can be rigorously tested,
 resulting in the approach we present.  
Similar ideas have been mentioned in passing previously by
Slowinski~\cite{Slowinski2001} and by Allman {\em et al.}~\cite{Allman} but to our knowledge, these suggestions have never been implemented or tested. 
The statistical test that we present is implemented inside the ASTRAL package
(option {\tt -t 10}) since version 4.11.2 and is available online at~\url{https://github.com/smirarab/ASTRAL}.

\section{Materials and Methods}

\subsection{Background}
An unrooted tree defined on a quartet of species $\{a,b,c,d\}$
can have one of three topologies (Fig.~\ref{fig:diagram}a): $t_1=ab\|cd$ (i.e., $a$ and $b$ are closer to each other than $c$, and $d$),
$t_2=ac\|bd$, or $t_3=ad\|bc$.
Consider an unrooted species tree $ab\|cd$
where the internal branch length separating species $a$ and $b$ from $c$ and $d$ is $x$ in coalescent units (CU), which is the number of generations divided by the haploid population size~\cite{degnan2009}. Under the Multi Species Coalescent (MSC) model, each gene tree
matches the species tree with the probability $p_1=1-\frac{2}{3}e^{-x}$ and matches each of the
two alternative topologies with the probability $p_2=p_3=\frac{1}{3}e^{-x}$~\citep{Allman}. 
Given true (i.e., error-free) gene trees
with no recombination within a locus but free
recombination across loci, 
frequencies $n_1,n_2,n_3$ of gene trees matching topologies $t_1,t_2,t_3$ will follow a
multinomial distribution with parameters
$p_1,p_2,p_3$ and with the number of trials equal
$n=n_1+n_2+n_3$.
Clearly, for a species tree with $N>4$ species,
the same results are applicable for any of the ${N \choose 4}$ selections of a quartet
of the species with $x$ defined
to be the branch length on the species
tree restricted to the quartet (see Fig.~\ref{fig:diagram} for examples). 

\subsection{A statistical test of polytomy}
A true polytomy is mathematically identical to a bifurcating node that has at least one adjacent branch with length zero;
the zero-length branch can be contracted in the binary tree to introduce the polytomy (e.g., compare branches P4 -- P6 in Fig.~\ref{fig:simpolyt}a to the multifurcating tree),
If the true species tree for a quartet of taxa
has a polytomy (i.e., $x=0$), then all gene tree topologies are equally likely with $p_1=p_2=p_3=\frac{1}{3}$. 
Thus, if we had the true $p_i$ values, we would
immediately know if the species tree has a polytomy.
However, we can never know true $p_i$ parameters;
instead, we have observations $n_1,n_2,n_3$ with
$\EX(n_i)/n=p_i$.
Luckily, multinomial distributions concentrate
around their mean.
As the number of genes increases, 
the probability of quartet frequencies deviating
from their mean rapidly drops;
for example, according to Hoeffding's inequality, the probability of divergence by $\epsilon$ drops
exponentially and  is no more than $2e^{-2\epsilon^2n}$.
This concentration gives us hope
that even though we never know true $p_i$ values
from limited data, we can design statistical
tests for a $p=\frac{1}{3}$ null hypothesis.
For an internal branch $\mathcal{B}$ in a 
bifurcating species tree, consider the following.

\begin{description}
\item[Null hypothesis:] The length of the internal branch $\mathcal{B}$ is zero; thus, the species tree
has a polytomy.
\end{description}

To test this null hypothesis, we can use
quartet gene tree frequencies
given three assumptions.
\renewcommand{\theenumi}{A\arabic{enumi}}%
\begin{enumerate}
\item All positive length branches in the given species tree are correct.
\item  Gene trees are a random error-free sample from 
the distribution defined by the MSC model.
\item We have $n\geq 10$ gene trees with at least a quartet relevant to $\mathcal{B}$.
\end{enumerate}
\renewcommand{\theenumi}{\arabic{enumi}}%
A1, which  we have previously
called the locality assumption~\cite{localpp-sm}, can be somewhat relaxed.
For each bipartition (i.e., branch) of the true
species tree, either that bipartition or one of its NNI rearrangements
should be present in the given
species tree.

We now describe 
expectations under the null hypothesis. 
With start with the $N=4$ case.
By the A2 assumption, frequencies $n_1,n_2,n_3$ follow a multinomial distribution with parameters $(p_1,p_2,p_3,n)$.
Under the null hypothesis $p_1=p_2=p_3=\frac{1}{3}$. Thus,under the null, 
\begin{equation}\label{eq:x2}
\chi^2=\frac{(n_1-n/3)^2}{n/3}+\frac{(n_2-n/3)^2}{n/3}+\frac{(n_3-n/3)^2}{n/3}
\end{equation}
is asymptotically a chi-squared random variable with 2 degrees of freedom~\citep{zar2007}. 
This chi-squared approximation for three equiprobable outcomes is a good approximation when $n\geq 10$~\citep{Koehler1980,zar2007,read2012}, hence our assumption A3. 
For smaller $n$'s an exact calculation of the critical value is required~\citep{read2012}, but we simply avoid applying our test for $n<10$.
Given the chi-squared random variable as
the test statistic, we can simply use a Pearson's goodness-of-fit statistical test. 
Thus, the $p$-value is the
area to the right of the $\chi^2$
test statistic (Eq.~\ref{eq:x2}) under the probability density function
of the chi-square distribution with 
two
degrees of freedom  (Fig.~\ref{fig:diagram}b).
This integral is available in various
software packages, including the Java package {\tt colt}~\cite{colt}, which we use.

With $N>4$, we apply the test described above to each branch
of the species tree independently.
For each branch $\mathcal{B}$, 
we will have multiple quartets around that branch. 
We say that a quartet  of species $\{a,b,c,d\}$
is  {\em around} the branch $\mathcal{B}$
when it is chosen as follows:
select an arbitrary leaf
$a$ from the subtree under the left child of $\mathcal{B}$,
$b$ from the subtree under the right child of $\mathcal{B}$,
$c$ from the subtree under the sister branch of $\mathcal{B}$,
and $d$ from the subtree under the sister
branch of the parent of $\mathcal{B}$ (this can be easily adopted for a branch
incident to root). 
Note that by assumption A1 (and its relaxed version),
the length of the internal branch
of the unrooted species tree induced down 
to a quartet around $\mathcal{B}$ is identical 
to the length of $\mathcal{B}$. 
Thus, under the null hypothesis, for any 
quartet around $\mathcal{B}$, we expect that
the length of the quartet branch should be zero. 
Thus,
any arbitrary 
selection of a quartet around the branch would enable us to use the same exact test we described before for $N=4$.

Following the approach we previously used for defining localPP~\cite{localpp-sm}, we can also use {\em all} quartets around the branch. 
More precisely, let $n_{i,j}$ for $1\leq i\leq 3, 1\leq j\leq n$
be the number of quartets around the branch $\mathcal{B}$ that in gene tree
$j$ have the topology $t_i$ and
let 
$f_{i,j}=\frac{n_{i,j}}{n_{1,j}+n_{2,j}+n_{3,j}}$. 
Then, we define 
\begin{equation}\label{eq:ni}
n_i=\sum_{j=1}^n f_{i,j} = \sum_{j=1}^n \frac{n_{i,j}}{n_{1,j}+n_{2,j}+n_{3,j}}\; .
\end{equation}
Given these $n_i$ values, we  use the $\chi^2$  test statistic
as defined by Equation~\ref{eq:x2}, just as before.

While we use Equation~\ref{eq:ni} mostly for computational
expediency, our approach can be justified. 
Let $x_{i,j,k}, 1\leq i\leq 3, 1\leq j\leq n, 1\leq k\leq m$  be an indicator variable
that is 1 if and only if the quartet
$k$ around the branch $\mathcal{B}$ has the topology $t_i$ in gene tree $j$.
Let $m_{i,k}=\sum_j x_{i,j,k}$
be
the number of gene trees where a quartet $k$  has the topology $t_i$.
Note that any quartet $k$
around $\mathcal{B}$ can be chosen;
thus, our hypothesis testing approach
would work if we define $n_i=m_{i,k}$ for {\em any} $k$
and use those $n_i$ values
in Equation~\ref{eq:x2}. 
In particular, 
the quartet with the median $m_{i,k}$ is
a valid and reasonable choice. 
Moreover, note that if all gene trees are complete, Equation~\ref{eq:ni} simplifies to
$n_i=\mean_k m_{i,k}$. 
We further assume that in the (unknown) distribution 
of $m_{i,k}$ values, the mean approximates the median. 
Thus, we approximate  $n_{i}=\mean_k (m_{i,k})\approx \median_k (m_{i,k})$.
We use this approximation because,
as it turns out, computing 
the mean is more computationally efficient than using the median. 

It may  initially  seem that computing the 
$n_i$ values requires computing $f_{i,j}$ values, which would require $O(N^4n)$ running time. 
This would be too slow for large datasets. 
Computing the median quartet score  
also requires $O(N^4n)$.
However, 
the mean quartet score 
can be computed efficiently in $O(N^2n)$
using the same algorithm that we have previously described for the localPP~\cite{localpp-sm}.
We avoid repeating the algorithm here
but note that it is based on 
a postorder traversal of each gene tree
and computing
the number of quartets shared between
the four sides of the branch $\mathcal{B}$
and each tripartition defined
by each node of each gene tree.
This traversal is adopted from
ASTRAL-II~\cite{astral2}.

When gene trees have missing data, the definition of $f_{i,j}$ naturally discards missing quartets. 
Similarly, if the gene tree $j$ includes a polytomy for a quartet, it is counted towards neither of the three $n_{i,j}$ values
 and so, is discarded. 
Then, Equation~\ref{eq:ni} effectively assigns a quartet $k$
missing/unresolved in a
gene tree $j$ to each quartet topology $i$ proportionally
to the number of present and resolved quartets in the gene tree $j$ with the topology $i$; in other words, 
a missing $x_{i,j,k}$ is imputed to ${\sum_k x_{i,j,k}}/{m_j}$ where $m_j$ is the number
of quartets present and resolved in the gene tree $j$. 
A final difficulty arises when {\em none} of the quartets defined around $\mathcal{B}$ are present or if all of the present ones  are unresolved in a multifurcating input gene tree. 
When this happens, we discard the gene tree for branch $\mathcal{B}$, reducing the number of genes $n$. Thus, the {\em effective} number of genes (i.e., effective $n$) can change from one branch to another based on patterns of gene tree taxon occupancy and resolution. 
Note that the  A3 assumption is with respect to this effective number of gene trees and not the total number. 

\subsection{Evaluations}

We examine the behavior of our proposed test, implemented
in ASTRAL 5.5.9,  on several simulated and empirical
datasets on conditions that potentially violate assumptions A1 and A2. 

The empirical datasets are 
a transcriptomic insect dataset~\cite{insects}, 
a genomic avian dataset~\cite{avian} with ``super'' gene trees resulting from statistical binning~\cite{binning}, 
two multi-loci {\em Xenoturbella} datasets by Rouse {\em et al.}~\cite{Rouse2016} and
Cannon {\em et al.}~\cite{Cannon2016}, and 
a transcriptomic plant dataset~\cite{1kp-pilot} (Table~\ref{table:biol-info}). 
Since in the empirical data, the true branch length
or whether a node should be a polytomy is not known, 
we will report the relationship between  the estimated branch lengths and $p$-values.
We will also randomly subsample gene trees to test how the amount of data impacts the ability to reject the null; for this, we focus on selected branches that have been difficult to resolve  in the literature.


\begin{table}[tb]
\caption{{\bf Datasets.} 
ref: the first publication to produce the dataset,
max height (known only for simulated dataset): the height of the tree in number of generations (population size fixed to $2\times 10^5$),  
$N$: number of species,
$n$: the maximum number of genes, 
$R$: number of replicates, 
$qs$: average ASTRAL quartet score as a measure of 
 gene tree discordance; computed  using the true species tree and true gene trees for simulated and the estimated species tree and estimated gene trees for the biological datasets, 
 $GE$: average distance between true and estimated gene trees (known for simulated dataset).}\label{table:sim-info}\label{table:biol-info}
\centering
\begin{tabular}{lllllllll}
\toprule
Type & Dataset & ref& max height	& $N$ & $n$ & $R$	& $qs$	& $GE$ \\
\midrule
\multirow{5}{*}{Biological}&Aves&\citep{avian} && 48 & 2022 & 1&0.64& \\
&insect&\citep{insects}&& 144  & 1478 & 1& 0.72& \\
&plant&\citep{1kp-pilot}& & 103& 844 & 1&0.89& \\
&{\em Xenoturbella}&\citep{Rouse2016} && 26 & 393 & 1&0.50& \\
&{\em Xenoturbella}&\citep{Cannon2016} & & 78 &  212  & 1&0.55& \\
\midrule
\multirow{5}{*}{Simulated}&S12A &new & 1.6M & 12 & 1000 & 50 & 0.82 & 36\%\\ 
&S12B &new & 1.6M & 12 & 1000 & 50 & 0.68 & 35\%\\
&{S201} &{~\citep{astral2}} & 10M	 & 201 & 1000 & 100 &0.94&25\%\\
&{S201}&{~\citep{astral2}}&2M	 & 201 & 1000 & 100 &0.72&31\%\\
&{S201}&{~\citep{astral2}}&500K & 201 & 1000 & 97 &0.49&47\%\\
\bottomrule
\end{tabular}
\end{table}

\begin{figure}[!htbp]
\centering
\includegraphics[width=0.95\textwidth]{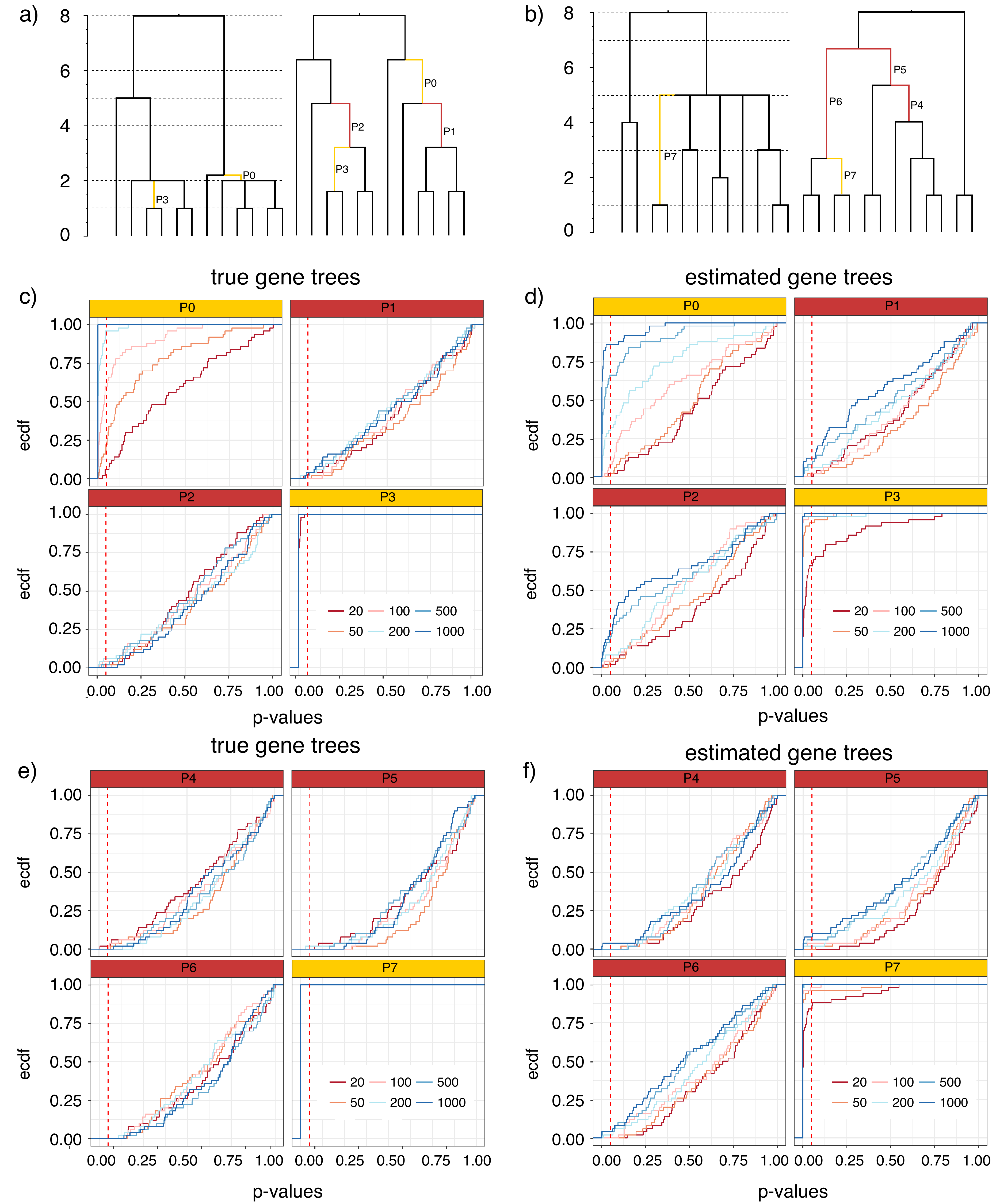}%
\caption{S12 datasets: true species trees and $p$-value distributions. For S12A (a) and S12B (b), we show the true multifurcating species trees in coalescent units (left) and an arbitrary resolution of the species tree used to test for polytomies  (right). 
Branches P1, P2, and P4--P6 (red) represent arbitrary resolutions for which the null hypothesis is correct. 
Branches P0, P3, and P7 (yellow) are selected as examples for which the null hypothesis is incorrect. 
(c--f) The $p$-value distributions are shown as empirical cumulative distribution functions (ECDF) where the x-axis shows the $p$-value $x$ and the y-axis shows the percentage of the replicates (out of 50) with a $p$-value $\leq x$.   
Results are shown for four selected branches of S12A (c,d)
and S12B (e,f) for both true gene trees (c,e) and estimated gene trees (d,f) with varying numbers of gene trees (line colors). 
Dashed vertical red line shows $p$-value$=0.05$. 
In red boxes, the intersection of the vertical line with each line shows the false positive rate.
In yellow boxes, the intersection of the vertical line with each line shows  one minus the false negative rate.
}
\label{fig:simpolyt}
\end{figure}

\paragraph{S12A and S12B} We simulated two datasets starting from two fixed species trees  with 12 species (S12A:  Fig~\ref{fig:simpolyt}a, S12B: Fig~\ref{fig:simpolyt}b). 
For both species trees,  the tree height is 1.6M generations  and the population size is $2\times 10^5$; thus, the tree height is 8 CU. 
The S12A species tree has two polytomies, each with three children, in addition to a short branch (P0) of length 0.2 coalescent units. The S12B tree has a polytomy with five children. 
For both S12A and S12b, Simphy~\cite{simphy} is used to simulate 50 replicates, each with 1000 gene trees. 
After generating the true gene trees, we used Indelible~\cite{Fletcher2009INDELible:Evolution} and the GTR+$\Gamma$ model of sequence evolution to simulate 250bp sequences down the gene trees. The GTR+$\Gamma$ parameters are drawn randomly from Dirichlet distributions used in the ASTRAL-II paper (parameters are estimated from a collection of biological datasets~\cite{astral2}).
We then used FastTree2~\cite{fasttree-2} to estimate gene trees from the sequence data. 
Both datasets have around $35\%$ gene tree error, measured as the average RF distance between true and estimated gene trees (Table~\ref{table:sim-info}).

On this datasets, we score an arbitrary resolution of the true multifurcating species trees. Therefore, we can have both false positive errors (incorrectly rejecting the null for a polytomy) and false negative errors (failing to reject the null for a positive-length branch).  
We vary the number of genes between 20 and 1000 by randomly subsampling them and examine the distribution of $p$-values across all 50 replicates for each interesting branch using both true and estimated gene trees. 

\paragraph{S201} 
We use a 201-taxon simulated dataset previously generated~\citep{astral2,localpp-sm}. Species trees are generated using the Yule process with a maximum tree height of 500K, 2M, or 10M generations
and speciation rates of  $10^{-6}$ (50 replicates per model condition) and  $10^{-7}$ (another
50 replicates).
The population size is fixed to $2\times 10^5$ in all datasets.
Thus, we have three conditions, each with 100 replicates and each tree includes 198 branches
(59,400 branches in total). 
Branch lengths have a wide range
(as we will see). 
The estimated gene trees on this dataset have relatively high levels
of gene tree error (Table~\ref{table:sim-info}).
Each replicates has 1000 gene trees, which we also randomly subsample to 50 and 200.

In this dataset, the true species trees are fully binary and therefore,
the null hypothesis is never correct. 
Any failure to reject the null hypothesis is  a false negative error. 
The inability to reject the null hypothesis should never be taken as accepting the null hypothesis because it can. 
simply indicate that the available data is insufficient to distinguish a polytomy from a short branch.
An ideal test should be able to reject the null for long branches. 
However, for very short branches, failing to reject the null  would be the expected behavior.
It is worth contemplating the meaning of super short branches.
For a haploid population size of $10^5$, a branch
length of $10^{-4}$ CU corresponds to only ten generations.
One can argue that such short branches, for most practical purposes,  can be considered a polytomy. 
Thus, false negative errors among super short branches could perhaps be tolerated.

\section{Results}

\subsection{Simulated datasets}

We focus our discussions on $\alpha=0.05$, but we show full distributions of $p$-values in many places. 

\paragraph{S12A and S12B} 


On the S12A tree, P1 and P2 are zero-length branches and therefore, the test should ideally fail to reject the null hypothesis for them.
As desired, when true gene trees are used, $p$-values are uniformly distributed (Fig.~\ref{fig:simpolyt}c; note the linear empirical cumulative distribution functions for P1 and P2 with true gene trees). For example, the null hypothesis is rejected  for 4\% of replicates with 1000 gene trees. 
As expected, since the null is correct, the false positive rate does not increase as we increase  the number of gene trees. 
Switching to estimated gene trees universally increases false positive errors (Fig~\ref{fig:simpolyt}d). For example for P1, we reject the null hypothesis in 12\% of replicates using 1000 gene trees.  The most severe case of false positive error rates occurs for branch P2, 
where  24\% of replicates are rejected with 1000 gene trees. 
Thus, gene tree errors can, in fact, increase the false positive error rates, but the extent of the increase depends on the length of branches surrounding the tested branch. 

On the S12A tree, we also examine two binary positive-length branches: P0, which is short (0.2 CU length) and the parent of a polytomy, and P3 (1 CU), which is longer and the child of a polytomy. On these,  we desire that the null hypothesis should get rejected.  
The P3 branch is easily rejected in all replicates using true gene trees.
With estimated gene trees, given 50 genes or more, the null is rejected in almost all cases, and is rejected in  66\% of replicates with 20 genes. 
Thus, the power to reject this moderate length branch (corresponding to $2\times 10^5$ generations) is very high. 
For  P0, which is rather short,  the ability to reject the null hypothesis depends on the number of genes and similar to other branches, the power is higher for true gene trees. 
The false negative rates decrease as the number of genes increases; using 1000 gene trees, the null is rejected in all replicates with true gene trees and in 86\% of replicates with estimated gene trees. 
Overall, the false negative rate is a function of the number of genes, 
the length of the branch, and gene tree error, as expected.

 
The S12B tree shows broadly similar results as S12A (Fig~\ref{fig:simpolyt}ef) but some differences are noteworthy. 
On the zero-length branches (P4, P5, and P6), as desired, the test fails to reject the null. 
However,  false positives rates are a bit lower than expected by chance when true gene trees are used. 
For example, at $\alpha=0.05$, we barely ever reject the null hypothesis for either of these three branches. 
These lower than expected false positive rates may be due to the fact that each branch is considered independently in our test, but P4, P5, and P6 are very much dependent (they all resolve one high degree polytomy). 
Even using estimated gene trees, the false positive rate remains low. 
With estimated gene trees, for P4, we reject the null in 4\% of replicates  when we use 1000 gene trees and we never reject the null hypothesis otherwise (Fig~\ref{fig:simpolyt}f). 
For P5 and P6, the false positive rates is at most 8\% and 4\% with 1000 genes.
While the false positive rates remain low with estimated gene trees,
the rate seems to slightly increase with increased numbers of gene trees. 
Alongside the zero-length branches, we also study the branch P7  (length: 2 CU), which is adjacent to the polytomy. 
For this relatively long branch, we always reject the null hypothesis with true gene trees. With estimated gene trees, the false negative rate is only 16\% with 20 gene trees and gradually drops to 0\% at 200 genes or more.

\begin{figure}[!tb]
\centering
\includegraphics[width=0.59\textwidth]{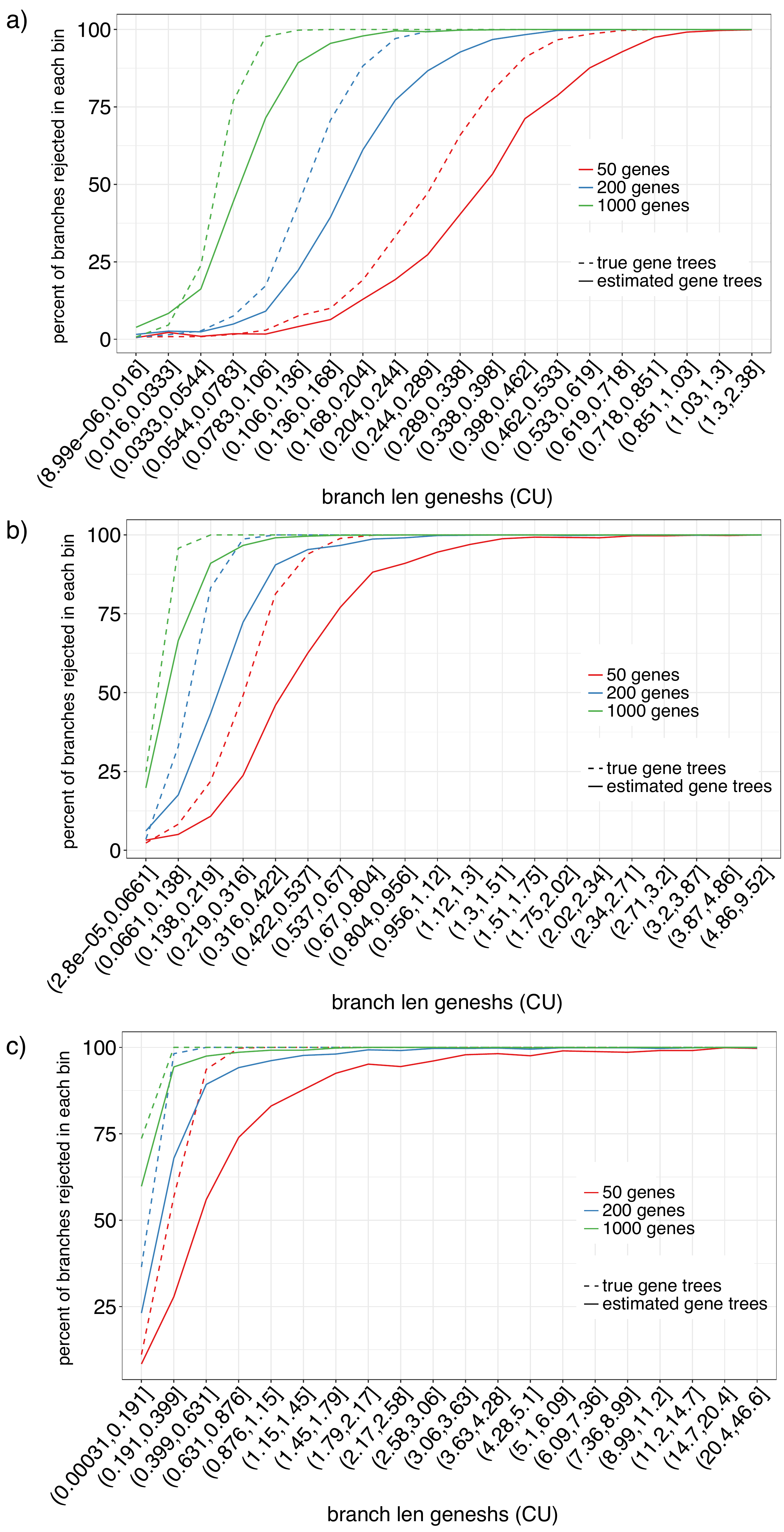}\\%
\caption{Polytomy test on S201 using estimated (solid) and true (dashed) gene trees for the different numbers of genes (colors)
for model conditions with the tree height set to
 500K  (a),  2M  (b), and  10M  (c) generations.
We show percentages of branches with the $p$-value $\leq 0.05$ (y-axis) for branch length ranges (x-axis), formed
by dividing the log of the true CU branch lengths into 20 equisized bins. }
\label{fig:simline}
\end{figure}

\paragraph{S201}
On the S201 datasets, we can only have false negative errors. 
We bin branches according to the log
of their CU length into 20 categories and compute the percentage of
branches that are rejected according to our
test with $\alpha=0.05$ per bin (Fig~\ref{fig:simline}). 
The false negative rate
mostly depends on three factors: 1) the branch length, 
2) the number of genes, and 3) whether true or estimated gene trees
are used.
The impact of all three factors is consistent with 
what one would expect for a reasonable statistical test.
For the longest branches (e.g., $>1.5$ CU), the null hypothesis
is rejected almost always even with as few as 50 genes
and with our highly error-prone estimated gene trees. 
Using the true gene trees instead of estimated gene trees
increases the power universally. 
For example, with 50 true gene trees, 
branches as short as $0.6$ CU are almost always rejected. 
Interestingly, the difference between estimated and
true gene trees seems to reduce as the number
of genes increase.

Reassuringly, as the number of genes increases, the power 
to reject the null hypothesis also increases. 
Thus, with 1000 genes, branches
between 0.1--0.2 CU are rejected 99.9\% of the times
with true gene trees
and 90.0\% of the times with estimated gene trees. 
Branches below $\log(7/6)\approx 0.15$ are considered very short and {\em can} produce the anomaly zone~\cite{Degnan2006,Degnan2013}.
Branches in the 0.05--0.15 CU range
are rejected 90.4\% and 67.4\% of times with 1000 true and estimated
gene trees, respectively.

\begin{figure}[!t]
\centering
\includegraphics[width=1.0\textwidth]{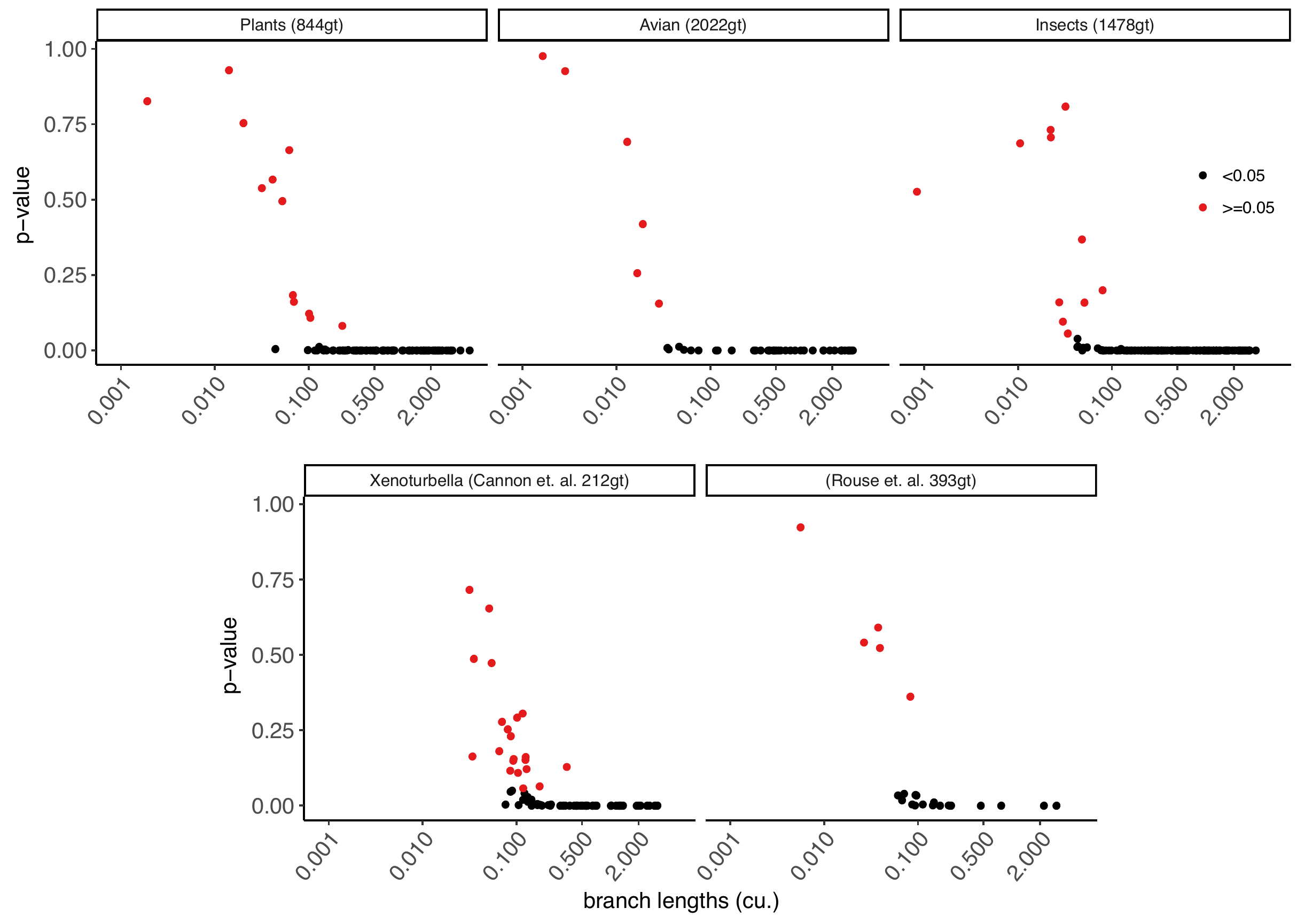}
\caption{Polytomy test results for 5 different biological datasets using ASTRAL species trees, and all available gene trees. For each internal branch, we show its ASTRAL estimated CU length  in log scale (x-axis) and its polytomy test $p$-value (y-axis). Points with $p<0.05$ are in black. For each dataset (panel), the number of genes is reported inside the parentheses in the title.  }
\label{fig:point-biolog}
\end{figure}

\subsection{Biological dataset}

On the biological datasets, 
the ability to reject the null hypothesis depends on the branch length and the effective number of gene trees (Figure~\ref{fig:point-biolog}). Most branches with the estimated length greater than $0.1$ CU had $p<0.05$. Datasets with more than a thousand genes (Aves and insects) had higher resolution and have $p<0.05$ for branches with the estimated length as lows as $0.035$~CU. 
Yet in all datasets except the Aves (where all gene trees include all species) there are some ranges of branch length (often above $0.1$ CU) where we are able to reject the null hypothesis for some branches but not for the others.
This cannot be just due to random noise because estimated (not the unknown true)
branch length is shown and two branches with the same length, 
have identical $n_i/n$ values. 
Instead, the reason is that the effective number of genes changes
from one branch to another because some gene trees may not include enough species to define a quartet around some branches.
The effective number of genes can also decline  due to a lack of gene tree resolution, but this does not happen in our datasets, which include only binary gene trees (we will revisit this in the discussion section).

\begin{figure}[!htb]
\centering
\includegraphics[width=0.74\textwidth]{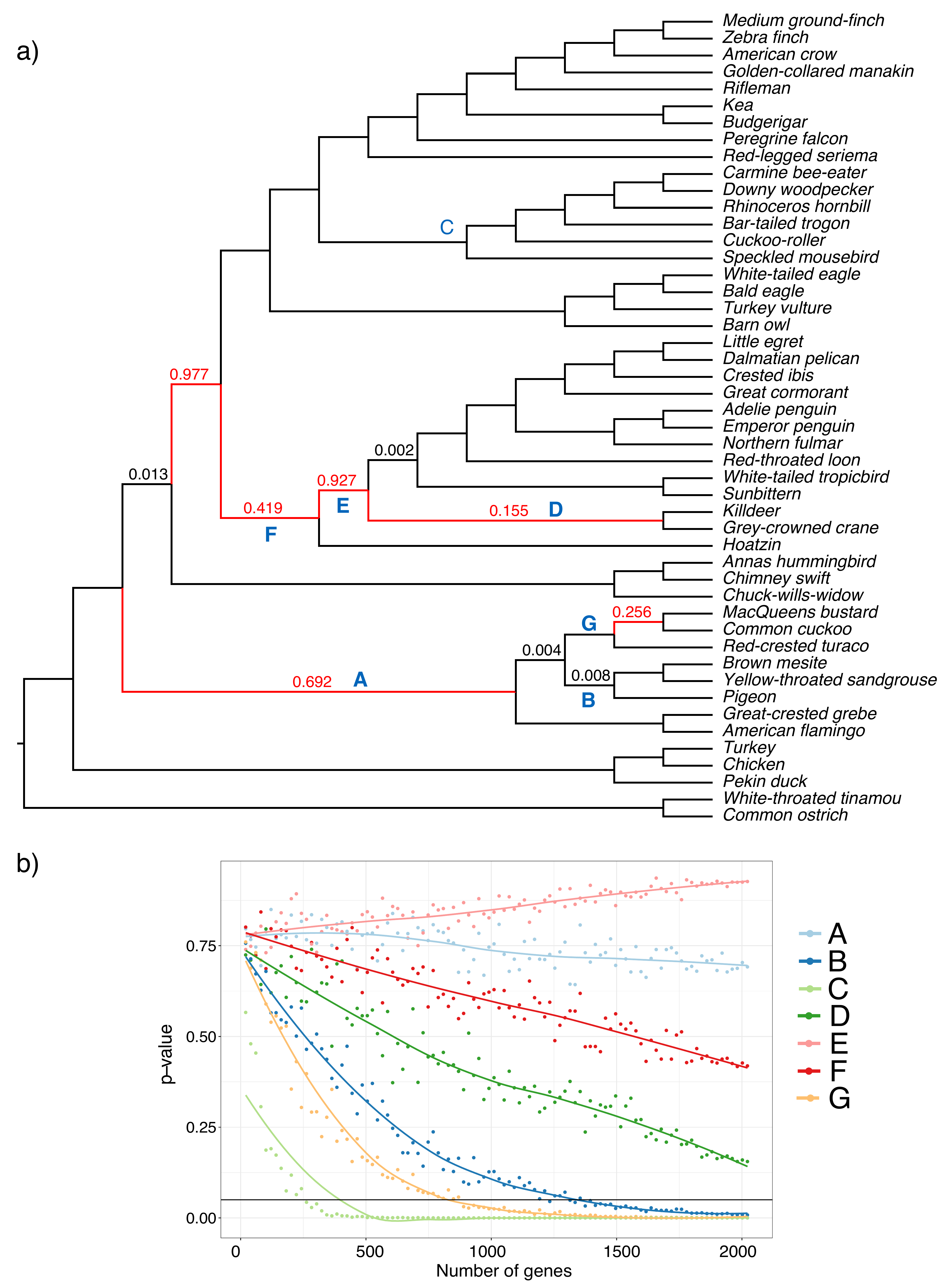}
\caption{{\bf Polytomy test results on avian dataset.} (a) ASTRAL species tree using binned ML gene trees. $p$-values greater than zero are reported on the branches, branches with $p>0.05$ are shown in red.
(b) change in $p$-value with respect to the number of genes for the selected branches in the species tree (labeled in blue in panel a). We used ASTRAL species trees with the varying number of gene trees sampled uniformly (1\%, 2\%, 3\%,\ldots, 100\% of gene trees but no less than 20), and repeated 20 times. We show average $p$-values (y-axis) versus the number of gene trees (y-axis). 
Solid horizontal line shows $p$-value$=0.05$. }
\label{fig:avian-biol}
\end{figure}

\begin{figure}[!htb]
\centering
\includegraphics[width=0.55\linewidth]{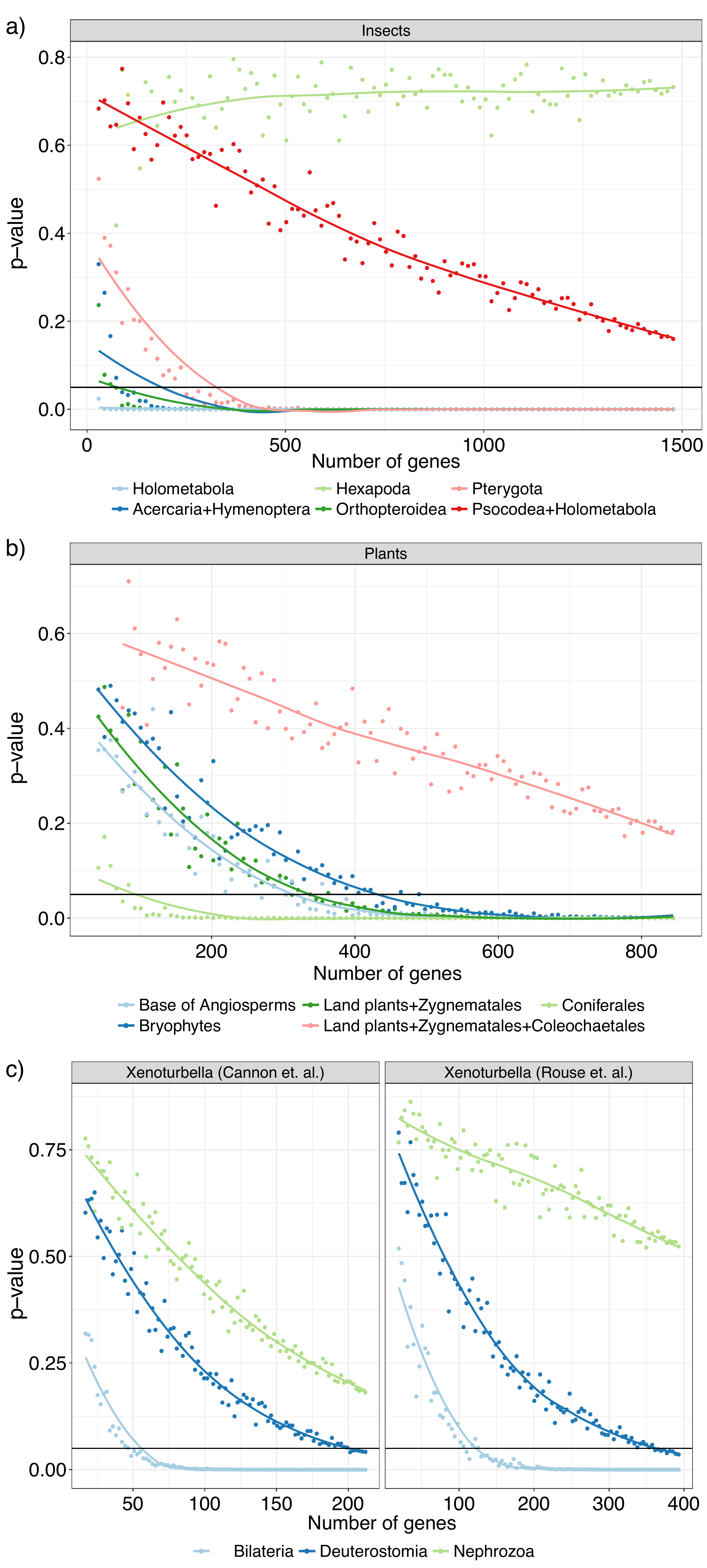}
\caption{Polytomy test results for selected branches of (a) insects (b) plants, and (c) two {\em Xenoturbella} datasets. 
We used ASTRAL species trees with the varying number of gene trees sampled uniformly at random (1\%, 2\%,\ldots, 100\% of gene trees but no less than 20) repeated 20 times. We show average $p$-values (y-axis) versus the number of gene trees (x-axis). 
Solid horizontal line shows $p$-value$=0.05$. Cases with effective $n$ below 10 are excluded; for plants and {\em Xenoturbella}, we omit 1--4\% because most replicates have $n<10$.}
\label{fig:biological}
\end{figure}

To further test the impact of the number of genes, for each dataset, we randomly subsampled gene trees ($1\%-100\%$ but no less than 20 gene trees) to find out how many genes are needed before we are able to reject the null hypothesis. 
We repeat this subsampling procedure 20 times, and show the average $p$-values across all 20 runs (Figures~\ref{fig:avian-biol} and~\ref{fig:biological}). 
In these analyses,  we focus on selected branches of each empirical dataset. 
Note that in some downsampled datasets, occasionally branches have an effective number of genes that is smaller than $10$, violating our assumption A3;  we exclude these branches.


For the avian datasets, 6 branches in the species tree could not be rejected as a polytomy at $\alpha=0.05$ even with all super gene trees (Figure~\ref{fig:avian-biol}a). 
These mostly belong to what has been called the wall-of-death~\cite{Joseph2015}, 
a hypothesized rapid radiation at the base of Neoaves~\cite{avian,Patel2013}.
In subsampling super gene trees, 
we highlight seven selected branches (labeled A--G) as shown in Figure~\ref{fig:avian-biol}a.
Interestingly, when we subsample super gene trees, several distinct patterns emerge for various branches.
Most branches are easily rejected as a polytomy even with a small fraction of the data (e.g., C). 
For some shorter branches (e.g., {G and B}) rejecting a polytomy requires hundreds of super gene trees. 
Yet for others (e.g., D and perhaps F), we cannot reject the polytomy with the full dataset, but the pattern suggests that if we had more super gene trees, 
we may have been able to reject them as a  polytomy.
Finally, for some branches (e.g., A and G), increasing the number of genes does  not lead to a substantial decrease in the $p$-value, suggesting that increasing the number of input trees may not be sufficient to resolve them. 

For the insect dataset we focus on 6 clades, Holometabola, Acercaria+Hymenoptera,  Hexapoda, Orthopteroidea, Pterygota, and Psocodea+Holometabola; these all have been classified as having {\em fairly strong support} from the literature~\cite{insects}, indicating that they enjoy robust support in the literature but some analyses reject them. 
As we reduce the number of genes, just like the avian dataset, we see three patterns (Figure~\ref{fig:biological}a).
For clades Holometabola, Acercaria+Hymenoptera, and Orthopteroidea, we get $p<0.05$ even with fewer than $100$ genes and for Pterygota with around 250 genes.
We are not able to reject the null hypothesis for  Psocodea+Holometabola with all gene trees, but the decreasing $p$-values suggest that this resolution could perhaps be resolved if we had several hundred more loci. 
The support for Hexapoda never decreases as we use more genes, suggesting that the relationship between insects and their close relatives (Collembola and Diplura, both considered insects in the past) may remain unresolved if we simply increase the number of genes.
For this deep (around 450M years old) and undersampled node,
 $p$-values may fail to reduce either
because of a true polytomy or because gene trees are estimated with high
(perhaps biased) error.

In remaining datasets (plants and {\em Xenoturbella}), all important branches that we studied  saw decreasing  $p$-values as the number of gene trees increase (Figure~\ref{fig:biological}b). 
In the plant dataset, having around 400 genes seems sufficient for most branches of interest, including the monophyly of Bryophytes and the resolution of Amborella as sister to all the remaining flowering plants.
The branch that puts Zygnematales as sister to land plants is rejected as a polytomy with about 350 genes.
However, the correct relationship between  Chara and Coleochaetales remains hard to resolve. Even with the full dataset, a polytomy is not rejected, though the decreasing $p$-values point to the possibility that this relationship would have been resolved had we had more genes. 

The {\em Xenoturbella} datasets both have three focal branches, surrounding the position of Xenacoelomorpha.
The
branch labeled Bilateria, which has Xenacoelomorpha and 
Nephrozoa as daughters branches in both papers,
can be resolved at $\alpha=0.05$ with as few as 50 (Cannon) or 100 (Rouse) gene trees (Figure~\ref{fig:biological}c). 
However, pinpointing the position of Xenacoelomorpha also depends
on the  branch labeled Nephrozoa, which puts Xenacoelomorpha
as sister to a clade containing Protostomia and Deuterostomia. 
The null hypothesis that this branch may be a polytomy is not rejected  in either dataset, 
but a pattern of decreasing $p$-values with more loci can be discerned. 
Thus, both datasets are best understood as leaving the relationships
between Protostomia, Xenacoelomorpha, and the rest of Deuterostomia as uncertain with some evidence that Xenacoelomorpha is at the base of Nephrozoa.
Remarkably, patterns of difficulty in resolving branches are similar across the two independent datasets with different taxon and gene selection.



\section{Discussion}

We introduced a new test for rejecting the null hypothesis that a branch in the species tree should be replaced by a polytomy. 
Unlike existing tests, our new test considers gene tree discordance due to ILS, as modeled by the MSC model. 
In several simulations, we showed that the test behaves as expected.
The null hypothesis is often retained for true polytomies and
is often rejected for binary nodes, unless when the true branch lengths are very short. 
The power to reject the null hypothesis for binary relationships increases with longer branches or with more gene trees and is reduced with gene tree estimation error. 
Gene tree error can also, in some cases, increase the false positive rate. 



\subsection{Power}
Overall, even when we have 1000 genes,
it is rare that we can reject the null for branches
shorter than 0.03 CU. 
A branch of length 0.03 corresponds to 6000
generations in our simulations. 
One can argue that failing to resolve a branch that corresponds to such short evolutionary times
(roughly 60K years with a generation time of 10 years)
can perhaps be tolerated. 
Mathematically, given a sufficiently large unbiased sample of gene trees, even infinitesimally short branches can be distinguished from a polytomy.
In practice, however, extremely short branches should be treated with suspicion as our input gene trees invariably are not perfect samples from the MSC distribution. 

In our biological analyses, we saw that subsampling genes and tracking trajectories of the $p$-value may be helpful in predicting the number of required genes to resolve a branch.
The approach we presented can be used in other biological data as well. However, we caution that such predictions should be interpreted with the limitations of our proposed test in mind. 
Many factors such as gene tree error and other sources of discordance can contribute to deviations from MSC, and such deviations may render the predictions inaccurate.  
But if such predictions are to be made, 
a natural question arises: How does the number of genes impact the  power?

\begin{figure}[t!]
\centering
\begin{scriptsize}
\hfill (a)\hfill (b)\hfill~\\
\includegraphics[width=0.45\linewidth]{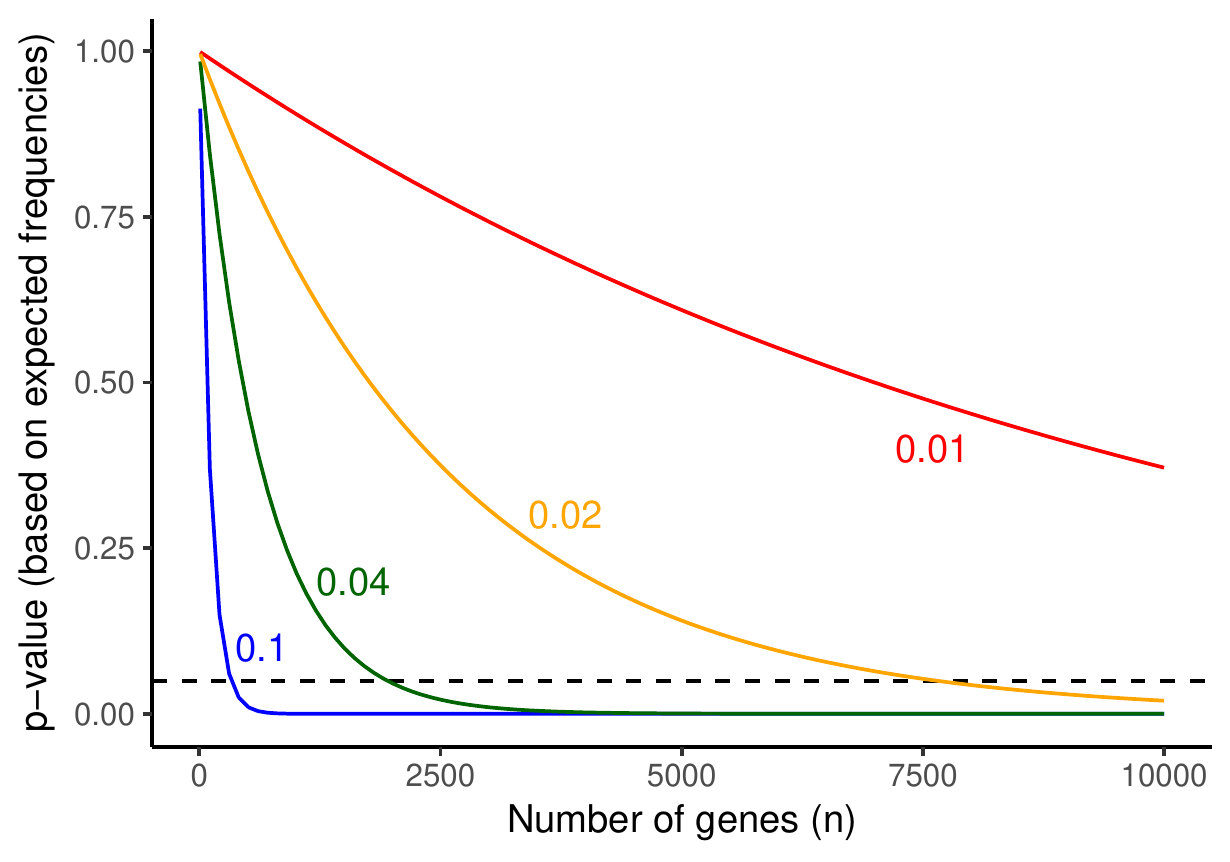}%
\includegraphics[width=0.45\linewidth]{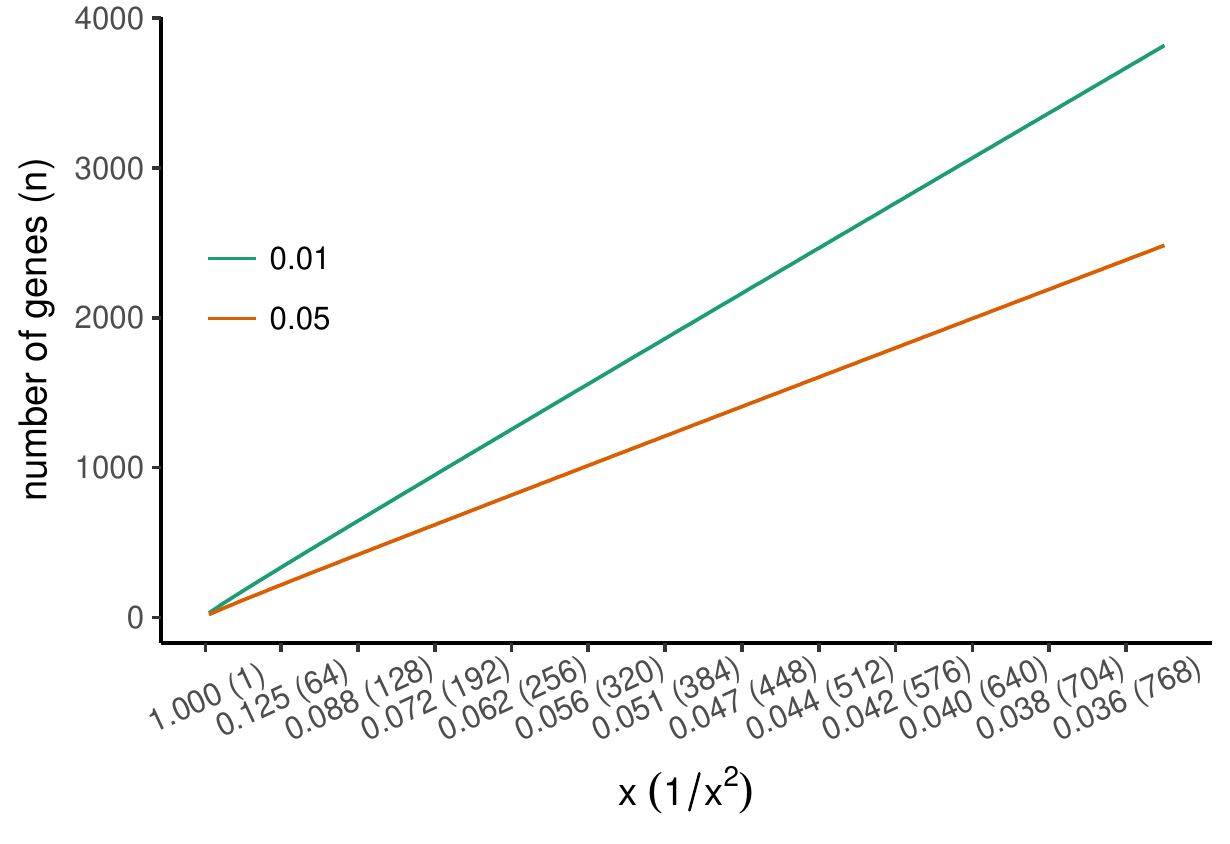}
\end{scriptsize}
\caption{{\bf Impact of the number of genes on $p$-value}
(a) The $p$-value computed for the different number of gene trees (x-axis) for four different
short branch lengths (colors) when the observed frequencies exactly match
the expected frequencies given that branch length. 
Dashed horizontal line shows $p$-value$=0.05$;
it intersects at 331 for 0.1 CU,
1949 for 0.04 CU, 7641 for 0.02 CU, and 30259 for 0.01 CU (not shown).
(b) The required number of genes (y-axis) to reject the null hypothesis with a $p$-value of 0.05 or 0.01
for various branch lengths (x-axis) assuming that the 
observed frequencies match the expected frequencies. 
Note that the x-axis scales with $\frac{1}{x^2}$.}
\label{fig:pvsn}
\end{figure}

We can easily compute  the required number of  genes for rejecting
the null hypothesis assuming the expected frequencies match observed
frequencies (Fig.~\ref{fig:pvsn}a).
For example, while for a branch of length 0.1 CU we only need $\approx$300 genes before we can reject it as a polytomy, for a branch of length 0.02 (i.e., 5 times shorter), we need $\approx$7500 genes (i.e., 25 times more).
For a quartet species tree, $n_1>\max (n_2,n_3)$
with arbitrarily high probability if the number
of genes grows as $\frac{\log N}{x^2}$~\cite{Shekhar2017}.
More broadly, the number of genes required for correct species tree estimation using ASTRAL is proven to grow proportionally to  $\log N$ and to $x^{-2}$~\cite{Shekhar2017}.
Similarly, for any given
branch length,  we can numerically compute  the minimum number of required
genes to obtain a given $p$-value (e.g., $0.05$).
Assuming the observed frequencies match the expectations,
we observe that the required number of genes grows 
linearly with $\frac{1}{x^2}$ (Fig.~\ref{fig:pvsn}b).
In fact, Figure~\ref{fig:pvsn}b gives us a way to estimate the level of ``resolution'' that a dataset can provide. For example, 300 genes can
reject the null for branches of $\approx$0.1 CU but if we quadruple the number of genes to 1200, our resolution is increased two-folds to
branches of $\approx$0.05 CU.
Note that gene tree error would increase these requirements and hence these should be treated as ballpark estimates. 
These estimates also assume we have $N=4$ species. 

The test we presented has no guarantees of maximal power. Other tests, such as likelihood ratio, may be more powerful. Moreover, it can be argued that our test is conservative in how it handles $N>4$. When multiple quartets are available around a branch, we use
their fraction supporting the $\mathcal{B}$ topology as the contribution of that gene to $n_1$. Thus, 
whether we have one quartet or a hundred quartets, we count each gene tree as one observation of our multinomial distribution. This is the most conservative approach to deal
with the unknown dependencies between quartets. 
The most liberal approach would 
consider quartets to be fully independent, increasing the degrees of the freedom of the chi-square distribution to $2m$ instead of $2$.
Such a test would be more powerful but would be based on invalid independence assumptions that may raise false positive rates. 
An ideal test would need to model the intricate dependence structure of quartets, a task that is very difficult~\cite{Erdos1999a}.

Finally, note that our test of polytomy relates to branch lengths in coalescent units. A branch of length zero in coalescent units will have length zero in the unit of time (or generations) if we keep the population size fixed. Mathematically, we can let
the population size grow infinitely. 
For a mathematical model where the population size grows asymptotically faster
than the time, one can have branches that converge to zero in length even though the branch length in time goes to infinity. 
This is just a mathematical construct with no biological meaning. Nevertheless, 
it helps to remind us that a very short branch in the coalescent unit (which our test may fail to reject as a polytomy) may be short not because the time was short but because the population size was large. 
Branches between 0.1 and 0.2 CU were not rejected as a polytomy by our test $\approx$10\% of times even with 1000 genes. A length of 0.1 CU {\em can} correspond to 10M generations if the haploid population size is 100M. 

\subsection{Divergence from the MSC model and connections to localPP}

\begin{figure}[bt!]
\centering
\includegraphics[width=0.8\linewidth]{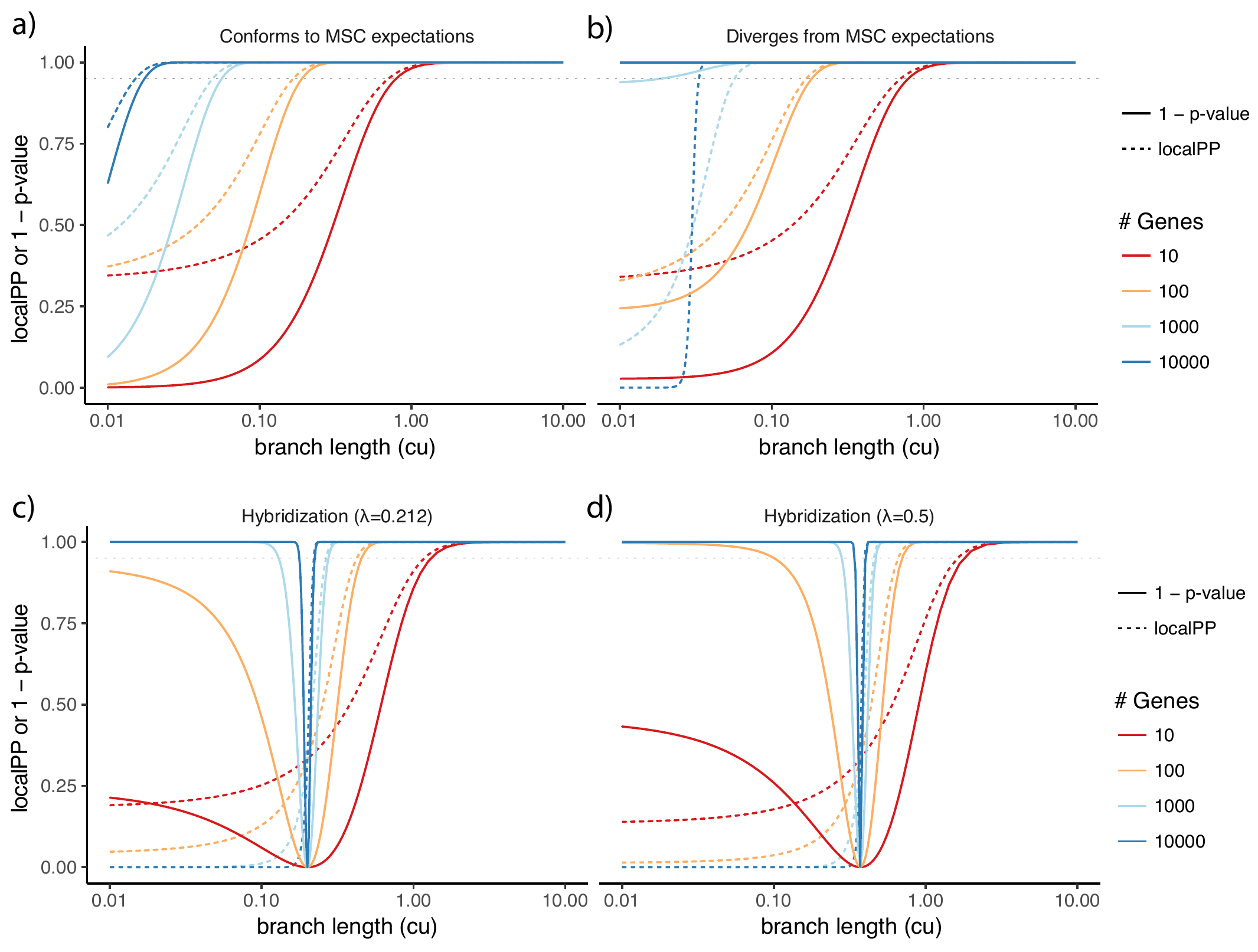}
\caption{{\bf The polytomy test versus localPP}.
For various branch lengths (x-axis; log scale) and various numbers of gene trees (colors), we show (y-axis) both
the localPP (dashed line) and
$1- p$-value of the polytomy test (solid line).
(a)
The quartet frequencies follow  MSC expectations:
$\frac{n_1}{n}=1-\frac{2}{3}e^{-x},\frac{n_2}{n}=\frac{n_3}{n}=\frac{1}{2}\frac{2}{3}e^{-x}$.
(b)
The quartet frequencies diverge
from the MSC expectations so that
$n_2$ is 20\% larger than $n_3$.
$\frac{n_1}{n}=1-\frac{2}{3}e^{-x}, \frac{n_2}{n}=\frac{6}{11}\frac{2}{3}e^{-x},\frac{n_3}{n}=\frac{5}{11}\frac{2}{3}e^{-x}$.
(c,d) quartet frequencies follow the MSC+gene flow model,
as analyzed by Solís-Lemus {\em et al.}~\cite{Solis-Lemus2016}.
For a species tree with a hybridization at the base (see Fig. 2 of~\cite{Solis-Lemus2016}) with 
inheritance probabilities $\lambda=0.1$ (c) and $\lambda=0.5$ 
(d), following
Solís-Lemus {\em et al.}, we set 
$\frac{n_1}{n}=(1-\lambda)^2(1-\frac{2}{3}e^{-x})+ 2\lambda(1-\lambda)(1-e^{-x/2}+\frac{1}{3}e^{-x-4})+\lambda^2(1-\frac{2}{3}e^{-x/2})$
and $n_2=n_3=\frac{n-n_1}{2}$.
The dotted horizontal gray line shows $p$-value$=0.05$. 
}
\label{fig:plocalPP}
\end{figure}

The  $p$-value from our proposed polytomy test has a close connection to the localPP branch support.
Both measures assume the MSC model and both are a function of quartet scores (i.e., $n_i/n$). 
As the quartet score of the species tree topology and the number of genes increases,
both localPP and $1-p$-value increase (Fig.~\ref{fig:plocalPP}a).
When localPP of a branch is close to 1.0,  the polytomy null hypothesis is always rejected.
However, the two measures are not identical.
Interestingly,
there are some conditions where localPP
is higher than 0.95 but the polytomy null hypothesis is not rejected at the 0.05 level (Fig.~\ref{fig:plocalPP}a).
When the frequencies follow expectations of the MSC model, $1-p$-value of the polytomy test is smaller than the localPP.

It is important to remember that our test relies on the properties of the MSC model. 
If observed quartet frequencies diverge from the expectations of the MSC model systematically (as opposed to by natural variation), the behavior of our proposed test can change. 
For example, if $n_2$ is substantially 
larger than $n_3$, rejecting the null hypothesis becomes easier (Fig.~\ref{fig:plocalPP}b). 
This should not come as a surprise because
this type of deviation from the MSC model makes the quartet frequencies even more diverged from $\frac{1}{3}$ than what is expected under the MSC model. 
On real data, several factors can may contribute to deviations from MSC. 
For example, incorrect homology detection in real datasets is possible (e.g., see~\cite{Springer2017} for possible homology issues with the avian dataset) and can lead to deviations.

Another source of deviation is gene flow, which can impact the gene tree distributions. Solís-Lemus {\em et al.} have identified {anomaly zone} conditions where the species tree topology has lower quartet frequencies compared
to the alternative topologies~\cite{Solis-Lemus2016}. 
Since the localPP measure does not model
gene flow, under those conditions, it will
be misled, giving low posterior probability to the species tree topology in the presence of gene flow (Fig.~\ref{fig:plocalPP}c). 
For example, if $\lambda=0.1$ (meaning that
10\% of genes are impacted by the horizontal gene flow), for branches of length 0.1 or shorter, localPP will be zero. 
The presence of the gene flow also impacts the test of the polytomy. 
For the species tree defined by Solís-Lemus {\em et al.} (Fig.~1 of ~\cite{Solis-Lemus2016}), 
when internal branches are short enough, 
there exist conditions where the gene flow and ILS combined result in quartet frequencies being equal to $\frac{1}{3}$ for all the three alternatives. 
It is clear that our test will not be able to distinguish such a scenario from a real polytomy (Fig.~\ref{fig:plocalPP}cd). 
One is tempted to argue that perhaps high levels of gene flow between sister branches {\em should} favor the outcome that the null is not rejected. However, 
this argument fails to explain 
the observation that for any value of
$\lambda$, the null hypothesis is retained
only with very specific settings of surrounding internal branch lengths (Fig.~\ref{fig:plocalPP}cd
).
Thus, we simply caution the reader
about the interpretation when gene flow 
and other sources of bias are suspected. 


\subsection{The effective number of gene trees}

\begin{figure}[!thb]
\centering
\includegraphics[width=0.95\linewidth]{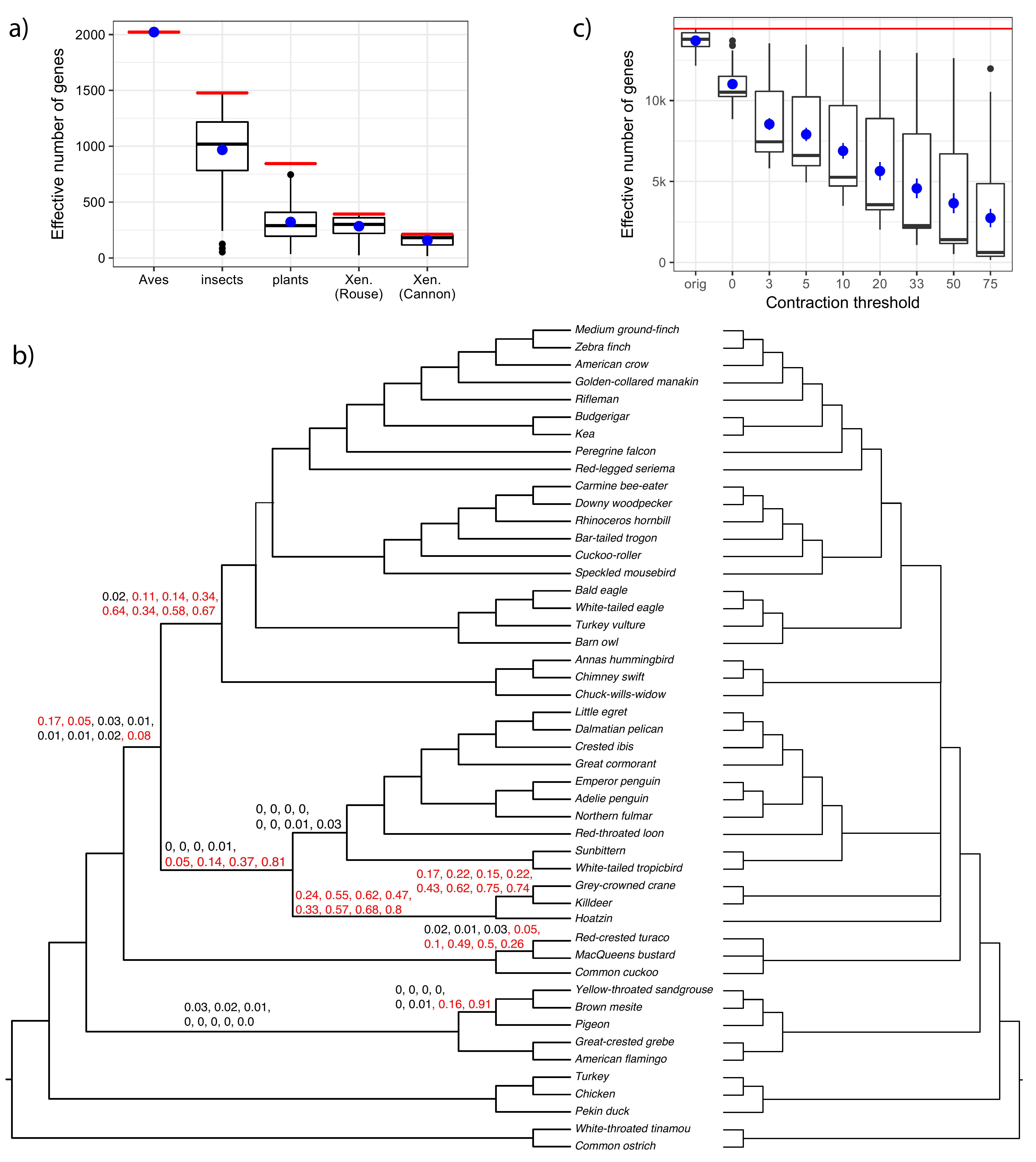}
\caption{{\bf Effective $n$ and results on the unbinned avian dataset}. (a) Distributions of effective $n$ (y-axis) across different branches of  each empirical dataset (x-axis). We show boxplots (black) as well as mean and standard error (blue). The total number of genes ($n$) is shown as a red horizontal line. 
(b) ASTRAL-III species tree estimated based on 14,446 unbinned gene trees with branches up to 10\% support contracted. 
For each branch, we show eight $p$-values that are computed, respectively, with respect to gene trees where branches with support up to 0\%, 3\%, 5\%, 10\% (top), 20\%, 33\%, 50\%, or 75\% (bottom) are contracted. 
Branches with no values have only 0 $p$-values (to three decimal points). 
$p$-values above 0.05 are in red. 
We also show the multifurcating species tree where all five branches that have $p$-values$<0.05$ according to the 10\% threshold  are contracted (the left facing tree).
(c) Similar to (a), we show distributions of effective $n$ (y-axis) across branches of the avian species tree with all 14,446 original unbinned trees (orig) 
or with gene tree branches with low support contracted (x-axis).}
\label{fig:effn}
\end{figure}

It is important to note that the effective number of gene trees (effective-$n$) can change across branches of the same species tree.  
Missing data can reduce the number of genes that have at least one taxon from a quartet defined around the branch of interest.  
In our biological datasets, various branches of the same dataset often have a wide range of effective $n$ (Fig~\ref{fig:effn}a), especially for the two transcriptomic datasets (insects and plants) with lots of missing data.
The only exception is the avian dataset, where our super gene trees always include all the taxa.

A second factor that can reduce the effective $n$ is multifurcations in input gene trees. If all the quartets around a branch are unresolved in an input gene tree, that gene tree does not count towards the effective $n$.  
Our biological datasets had binary gene trees. 
However, as recently shown~\cite{astral3}, removing branches with very low support can help addressing gene tree error. 
To demonstrate this, we revisit the avian dataset. 
The purpose of using super gene trees instead of normal (unbinned) gene trees was to reduce the gene tree estimation error.
Alternatively, one can simply remove branches with support at or below a certain threshold in gene trees and use the resulting tree as input to ASTRAL~\cite{astral3}.
With this procedure and the support threshold set to 10\%, we generated a new ASTRAL tree based on all 14,446 unbinned gene trees from the avian dataset~\cite{avian,binning} (Fig~\ref{fig:effn}b).
The resulting tree was largely congruent with the ASTRAL tree on super gene trees and with reference phylogenies form the original publication~\cite{avian}. 

We tested how the effective $n$ and $p$-values change as a result of contracting low support branches. 
Simply contracting branches with 0\% support reduces the median effective $n$ from 13,791 to 10,523. Further contracting branches with support up to 3\% -- 75\% gradually reduces the effective $n$ all the way to a median of 610 (Fig~\ref{fig:effn}c).
The $p$-values tend to decrease as we increase the threshold for contraction (Fig~\ref{fig:effn}b). 
Several branches fail to reject the null hypothesis regardless of the threshold chosen. 
Others reject the null hypothesis with lower levels of contraction but not with the higher levels, showing that the reduced effective $n$ can reduce the power. 
For one branch, interestingly, the null is not rejected if we contract up to 0\% and 3\% support {\em or} if we contract up to 75\%, but is rejected otherwise. 
This pattern may have a subtle explanation. 
With gene trees that include low support branches (up to 3\%), we are unable to reject the null hypothesis perhaps because gene tree error creates a uniform distribution of quartets around this branch.
 As we further remove low support branches from the gene trees, we start to see quartet frequencies that favor the ASTRAL resolution perhaps because noise is removed and the actual signal can be discerned. 
 Finally, with aggressive filtering of gene tree branches, effective $n$ becomes so low 
 that the test simply does not have the  power to reject the null. 
 These interesting patterns suggest that dealing with gene tree error by contracting  low support branches may be possible, but the choice of the best threshold is not obvious. 
 Future studies should further consider this question.

\subsection{Interpretation}
In the light of the dependence of our test on the MSC properties, we offer an alternative description of the test. 
A safe way to interpret the results of the test, regardless of the causes of gene tree discordance, is to formulate the null hypothesis as follows. 
\begin{description}
\item[Null hypothesis:] The estimated gene tree quartets around the branch $\mathcal{B}$ support all three NNI rearrangements around the branch in equal numbers.
\end{description}
This is the actual null hypothesis that we test. Under our assumptions, this hypothesis is equivalent to branch $\mathcal{B}$ being a polytomy. 
Under more complex models, such as gene flow + ILS, this null hypothesis holds true
for polytomies but also for some binary networks.

The judicious application of our test will preselect the branches where a polytomy null hypothesis is tested and examines the $p$-value only for those branches. 
When many branches are tested, one arguably needs to correct for multiple hypothesis testing, further reducing the power of the test. Corrections such as Bonferroni or FDR~\cite{Benjamini1995} can be employed (but we did not apply them in our large scale tests that did not target specific hypotheses). 
However, note that even though we formulate the polytomy as a null hypothesis, in reality, we expect that in most cases the branch has positive branch length. 
Thus, we expect to reject the null often, in contrast to usual applications of the frequentist test. 
The analyst should specify in advance the branches for which a polytomy null hypothesis is reasonable. 
This adds subjectivity, but
such problems are always encountered with
frequentist tests, and ours is no exception. Our test also suffers from
all the various criticisms leveled against the frequentist hypothesis testing~\cite{Anderson2000} and the interpretation has to avoid all the common pitfalls~\cite{Goodman2008}.

\section{Conclusions}
We presented a statistical test, implemented in ASTRAL, for the null hypothesis that a branch of a species
tree is a polytomy given a set of gene trees. Our test, which relies
on the properties of the multi-species coalescent model, performed well on  simulated and real data. As expected, its power
was a function of branch length, the number of genes, and the 
gene tree estimation error.

\vspace{6pt} 

\section*{supplementary}
{All data, R script, and results are available  at~\url{https://github.com/esayyari/polytomytest}.}

\section*{Acknowledgements}
This work was supported by the National Science Foundation grant IIS- 1565862 to E.S. and S.M. Computations were performed on the San Diego Supercomputer Center (SDSC) through XSEDE allocations, which is supported by the NSF grant ACI- 1053575.



\section*{Abbreviations}
The following abbreviations are used in this manuscript:\\

\noindent ILS: Incomplete Lineage Sorting\\
MSC: Multi-Species Coalescent\\
CU: Coalescent Unit



\bibliographystyle{mdpi}

\renewcommand\bibname{References}

\bibliography{polytomytest}


\end{document}